\documentclass[prb,showpacs,preprint,floatfix]{revtex4}

\usepackage{graphicx}
\begin{document}
\bibliographystyle{apsrev}
\title{The S=1/2 chain in a staggered field: High-energy bound-spinon
state \\ and the effects of a discrete lattice}

\author{M. Kenzelmann$^{1,2}$, C.~D. Batista$^{3}$, Y. Chen$^{1}$,\\
C. Broholm$^{1,2}$, D.~H. Reich,$^{1}$ S. Park$^{2,4,5}$ and Y.
Qiu$^{2,4}$ }

\affiliation{(1) Department of Physics and Astronomy, Johns Hopkins
University, Baltimore, MD 21218\\(2) NIST Center for Neutron
Research, National Institute of Standards and Technology,
Gaithersburg, MD 20899\\(3) Center for Nonlinear Studies and
Theoretical Division, Los Alamos National Laboratory, Los Alamos, NM
87545\\%
(4) Department of Materials Science and Engineering, University of
Maryland, College Park, MD 20742\\%
(5) HANARO Center, Korea Atomic
Energy Research Institute, Daejeon, Korea}
\date{\today}

\begin{abstract}
We report an experimental and theoretical study of the
antiferromagnetic $S$=$\frac{1}{2}$ chain subject to uniform and
staggered fields. Using inelastic neutron scattering, we observe a
novel bound-spinon state at high energies in the linear chain
compound ${\rm CuCl_2 \cdot 2((CD_3)_2SO)}$. The excitation is
explained with a mean-field theory of interacting $S$=$\frac{1}{2}$
fermions and arises from the opening of a gap at the Fermi surface
due to confining spinon interactions. The mean-field model also
describes the wave-vector dependence of the bound-spinon states,
particularly in regions where effects of the discrete lattice are
important. We calculate the dynamic structure factor using exact
diagonalization of finite length chains, obtaining excellent
agreement with the experiments.
\end{abstract}
 \maketitle

\section{Introduction}

When quantum particles interact, they can team up to form new
particles with fractional charge or spin as observed in
two-dimensional electron gases or low-dimensional spin lattices. A
particularly fruitful model system to study emerging new particles
and their interactions has been the antiferromagnetic (AF)
$S$=$\frac{1}{2}$ Heisenberg chain. It does not order even at
$T=0\;\mathrm{K}$ due to strong quantum fluctuations and its
elementary excitations are spinons carrying fractional
$S$=$\frac{1}{2}$, which interact only weakly and are unbound
particles.\cite{Faddeev_Takhtajan} Spinons were observed in several
materials, including ${\rm KCuF_3}$ \cite{Tennant95}, ${\rm
BaCu_2Si_2O_7}$\cite{Kenzelmann_BaCu2Si2O7} and copper pyrazine
dinitrate.\cite{Stone} Strong interactions between spinons can arise
from the breaking of spin rotational symmetry as for example in the
three-dimensional coupled chain antiferromagnet where the mean field
associated with the long-range ordered ground state restricts spin
fluctuations. This creates an attractive potential for the spinons
and at low energies they condense into spin-wave excitations
carrying $S$=$1$,\cite{Schulz96} as observed in ${\rm KCuF_3}$
\cite{Tennant95/2} and ${\rm
BaCu_2Si_2O_7}$.\cite{Zheludev_BaCu2Si2O7}\par

Recently excitations were observed in the $S$=$\frac{1}{2}$ chain
antiferromagnet ${\rm CuCl_2 \cdot 2((CD_3)_2SO)}$ (CDC) subject to
uniform and staggered magnetic fields which were interpreted as
bound spinon states.\cite{Kenzelmann_CDC_PRL} In the
long-wave-length limit, the experiment demonstrated that the
low-energy excitations of a $S$=$\frac{1}{2}$ chain in a staggered
field correspond to the soliton and breather excitations of the
quantum sine-Gordon model.\cite{Affleck_Oshikawa} However, the
sine-Gordon model is only valid for a very restricted range of
wave-vectors and does not apply for excitations at smaller length
scales where the discreteness of the lattice becomes important. More
importantly, it does not fully describe the mechanism by which the
staggered field produces an attractive potential that binds spinons
into the long-lived dispersive $S$=$1$ excitations as observed in
the experiment.\par

In this paper, we analyze the dispersion of the observed bound
spinon states \cite{Kenzelmann_CDC_PRL} in detail, and we report the
observation of a bound-spinon state at high energies in CDC. In this
system which has a nearest neighbor exchange $J=1.5\;\mathrm{meV}$,
a staggered field of the order of $1\;\mathrm{T}$, which corresponds
to a Zeeman energy $g\mu_B H \sim 0.1\;\mathrm{meV}$, qualitatively
affects the excitation spectrum to energies more than twice $J$ at
$3.4\;\mathrm{meV}$. This is in stark contrast to spinon binding in
coupled chain magnets where the effects are only apparent for $\hbar
\omega \approx k_B T_N$.\cite{Zheludev_BaCu2Si2O7} A simple
mean-field theory of fermions carrying $S$=$\frac{1}{2}$ in one
dimension captures the wave-vector dependence of the bound spinon
states, and explains the high-energy excitation through the opening
of a gap at the Fermi surface. This model represents a first step
towards a comprehensive description of the incommensurate
excitations in AF $S$=$\frac{1}{2}$ chains subject to staggered
fields.\par

\section{Experimental}

CDC was identified as an AF $S$=$\frac{1}{2}$ chain system in which
a staggered $g$-tensor and/or Dzyaloshinskii-Moriya (DM)
interactions \cite{Landee,Chen_CDC} lead to a staggered field
$H_{\rm st}$ upon application of a uniform field $H$. In CDC, a
uniform magnetic field ${\bf H}=(0,0,H)$ along the c-axis generates
a staggered field ${\bf H}_{\rm st}=(H_{\rm st},0,0)$ along the
a-axis, and the Hamiltonian can be written as
\begin{equation}
    \mathcal{H}=\sum_i J {\bf S}_i\, {\bf S}_{i+1} - g_c \mu_B H
    S^z_i - g_a\mu_B H_{\rm st} (-1)^i S^x_i\, ,
\end{equation}where $g_c$ and $g_a$ are the uniform part of the
gyromagnetic tensor along the c and a-axis, respectively. The
staggered field is given by
\begin{equation}
    H_{\rm st}=\frac{1}{2J} \frac{g_c}{g_a}D H + \frac{g_s}{g_a} H\,
    ,
\end{equation}where $g_s$ is the staggered gyromagnetic form factor
and $D=|{\bf D}|$ is the length of the DM vector, which points along
the b-axis in CDC. The nearest-neighbor spin exchange along the
chain is accurately known from susceptibility measurements
\cite{Landee} and inelastic neutron
scattering.\cite{Kenzelmann_CDC_PRL} The spin chains run along the
${\bf a}$-axis of the orthorhombic crystal structure ({\it
Pnma}),\cite{Willett_Chang} with the ${\rm Cu^{2+}}$ ions separated
by $0.5{\bf a} \pm 0.22{\bf c}$. Wave vector transfer is indexed in
the corresponding reciprocal lattice ${\bf Q}(hkl)=h{\bf a}^*+k{\bf
b}^*+l{\bf c}^*$, and we define the wave-vector transfer along the
chain as $q={\bf Q}\cdot {\bf a}$. Due to weak inter-chain
interactions, CDC has long-range AF order in zero field below $T_N =
0.93\;\mathrm{K}$ with an AF wave-vector ${\bf Q}_m={\bf a}^*$. An
applied field along the c-axis suppresses the ordered phase in a
second order phase transition at $H_c=3.9\mathrm{T}$,\cite{Chen_CDC}
indicating that inter-chain interactions favor correlations that are
incompatible with the field-induced staggered
magnetization.\cite{Oshikawanew} At fields much greater than $H_c$,
the staggered fields thus arise mostly from a staggered $g$-tensor
and DM interactions and not from interchain interactions.\par

The neutron scattering experiments were performed on
$7.76\;\mathrm{g}$ of deuterated single-crystalline CDC. The
measurements were carried out using the SPINS triple axis
spectrometer and the DCS time of flight spectrometer at the NIST
Center for Neutron Research. The SPINS measurements were performed
with a focussing analyzer covering $7^{\rm o}$ in $2\Theta$
scattering angle set to reflect $E_f=5\;\mathrm{meV}$ to the center
of the detector. A Be filter rejected neutrons with energies higher
than $5\;\mathrm{meV}$ from the detection system. The measurements
were performed with the strongly dispersive direction along the
scattered neutron direction to integrate over wave-vectors along a
weakly-dispersive directions. The experimental configuration for the
measurements made using the DCS spectrometer and the conversion of
those data to absolute units are described in detail
elsewhere.\cite{Kenzelmann_CDC_PRL} They were performed with an
incident energy $E_i$=$3.03\;\mathrm{meV}$ and the incident beam
parallel to the a-axis for configuration A, and with an incident
energy $E_i$=$4.64\;\mathrm{meV}$ and angle of $60^{\rm o}$ between
the incident beam and the a-axis for configuration B.\par

\begin{figure}[ht]
\begin{center}
  \includegraphics[height=6cm,bbllx=70,bblly=250,bburx=491,
  bbury=590,angle=0,clip=]{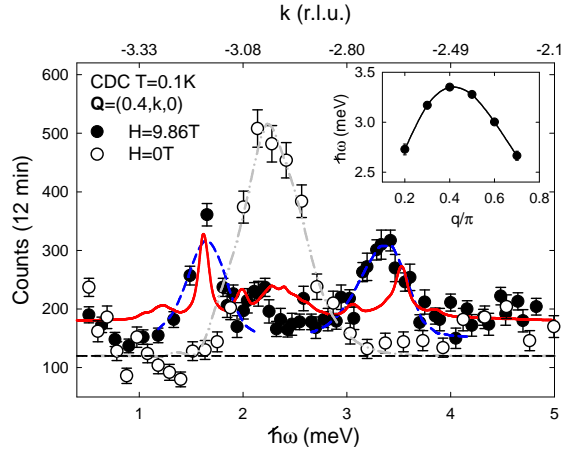}
  \caption{Neutron scattering intensity as a function of energy transfer
  for zero-field and $9.86\;\mathrm{T}$ measured using SPINS. The dashed
  line is a fit of two Gaussians convolved with the resolution function
  given by Cooper and Nathans.\protect\cite{Cooper_Nathans} The solid
  line shows the calculated intensity obtained from the exact
  diagonalization of finite chains for $H=11\;\mathrm{T}$ and scaled to
  the data, with the non-magnetic background given as the straight
  dashed line.  The dashed-dotted line is the exact two-spinon
  cross-section of antiferromagnetic $S$=$\frac{1}{2}$ chain
  \protect \cite{Bougourzi_Karbach} convolved with the experimental
  resolution function. Inset: Excitation energy of the higher-energy
  mode for $H$=$9.86\;\mathrm{T}$ as a function of wave-vector transfer
  along the chain direction.}
  \label{Fignewexcitation}
\end{center}
\end{figure}

\section{Experimental Results}

Figure~\ref{Fignewexcitation} demonstrates the dramatic changes that
CDC undergoes upon application of a magnetic field. In zero field,
the neutron scattering spectrum  for $q=0.4\pi$ consists of a strong
peak which corresponds to the two-spinon continuum, whose band
width, at this wave vector is narrow and barely distinguishable from
the experimental resolution. In a $H$=$9.86\;\mathrm{T} \simeq
\frac{3}{4} J/(g\mu_B)$ field applied along the c-axis, which also
induces a staggered field along the a-axis, the scattering includes
two resolution-limited excitations. According to the quantum
sine-Gordon theory, the lower-energy excitation corresponds to a
bound-spinon state which develops into solitons and breathers at
long wave-lengths. The high-energy excitation at $3.4\;\mathrm{meV}
\simeq 2.2 J$, however, does not have a simple interpretation in
terms of the sine-Gordon model. Its magnetic nature is apparent on
account of its field dependence. The inset to
Fig.~\ref{Fignewexcitation} shows the dispersion of this excitation,
which has a maximum at $q$=$0.4\pi$.\par

\begin{figure}[ht]
\begin{center}
  \includegraphics[height=10cm,bbllx=70,bblly=136,bburx=500,
  bbury=690,angle=0,clip=]{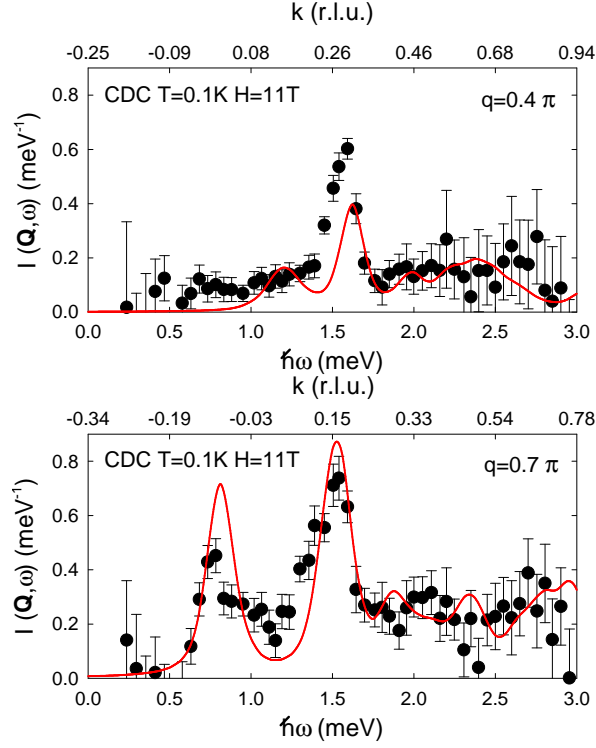}
  \caption{Neutron scattering intensity as a function of energy
  transfer at $11\;\mathrm{T}$ for two different chain wave vectors,
  measured using the DCS instrument with an incident energy
  $E_i$=$4.64\;\mathrm{meV}$. The solid line is the spectrum
  calculated from exact diagonalization of finite chains in absolute
  units, taking into account the polarization dependence of the
  experiment.}
  \label{Figotherscans}
\end{center}
\end{figure}

\begin{figure}[ht]
\begin{center}
  \includegraphics[height=11cm,bbllx=75,bblly=68,bburx=500,
  bbury=570,angle=0,clip=]{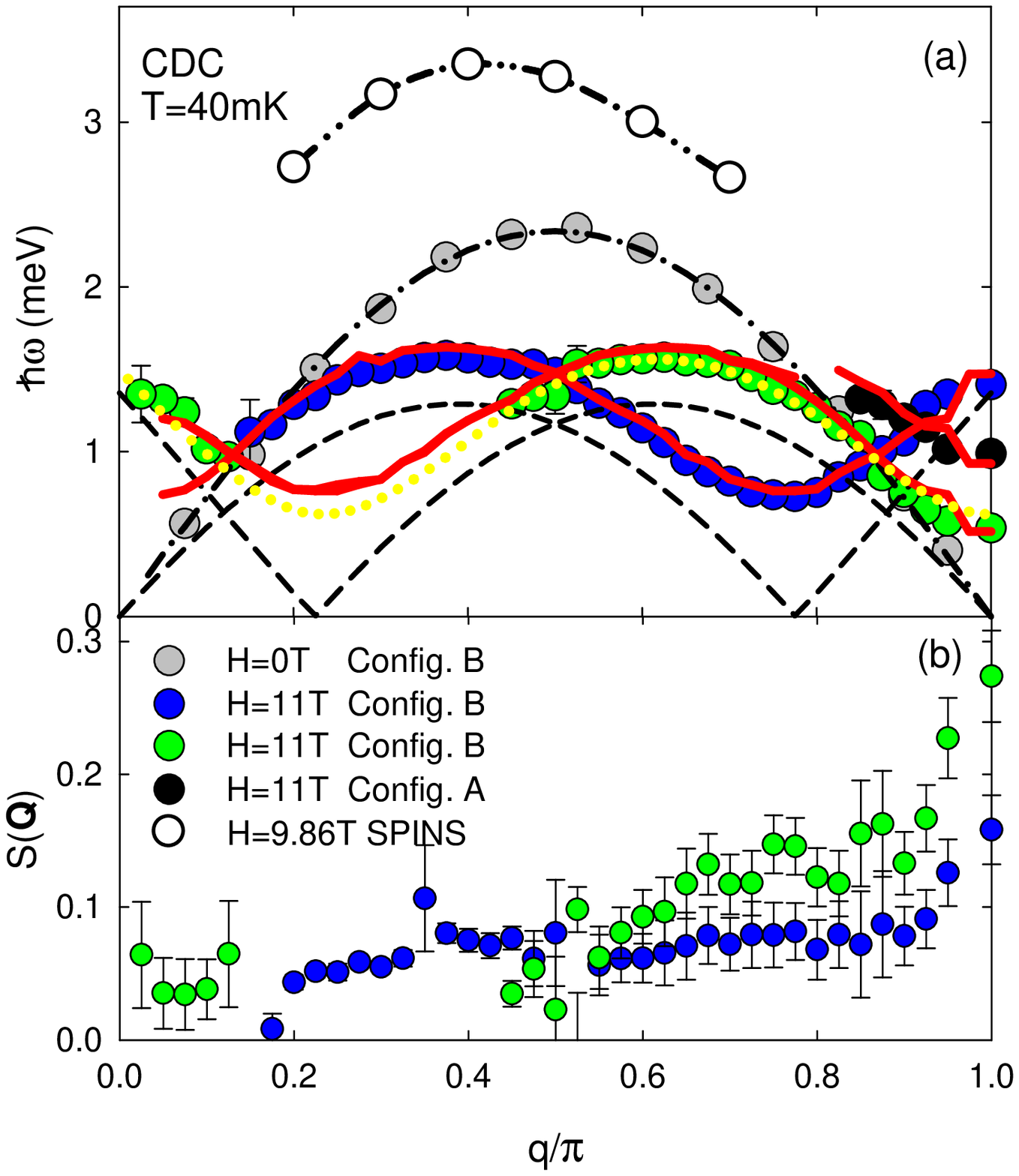}
  \caption{(a) Dispersion of the excitations at
  $H$=$11\;\mathrm{T}$ and the lower bound of the zero-field
  two-spinon cross-section,\protect\cite{Karbach_Bougourzi}
  obtained through fits to the zero-field data. The broken lines
  are predicted thresholds for continua for a $S$=$\frac{1}{2}$ chain
  in a uniform field.\protect\cite{Muller} The dashed-dotted line is
  the des-Cloizeaux-Pearson lower bound for excitations in
  $S$=$\frac{1}{2}$ chains in zero field \protect\cite{Cloizeaux_Pearson}
  for $J$=$1.5\;\mathrm{meV}$. The solid lines correspond to the
  dispersion obtained from Gaussian fits to the spectra obtained by
  the exact diagonalization of finite chains. The dotted line
  corresponds to the mean-field dispersion. The dashed double-dot
  line is a guide to the eye for the high-energy mode above
  $3\;\mathrm{meV}$, which was measured at $9.86\;\mathrm{T}$. (b)
  Integrated intensity of the two resonant modes at
  $H$=$11\;\mathrm{T}$ as a function of
  wave-vector transfer.}
  \label{Figdispersion}
\end{center}
\end{figure}

The dispersion of the high-field excitations at lower energies,
$\hbar\omega < 2\;\mathrm{meV}$, is illustrated in
Fig.~\ref{Figotherscans} for two different chain wave-vectors $q$.
For $q$=$0.7\pi$, there are two maxima as a function of energy,
corresponding to the well-defined modes developing into the
sine-Gordon soliton and breather excitations at long wave-lengths.
For $q$=$0.4\pi$, only one peak is clearly observed because of the
weak intensity of one of the modes in this wave-vector region. The
dispersion of these field-induced resonant modes was determined as a
function of the chain wave-vector, $q$, by fitting
resolution-corrected Gaussian line-shapes to the observed
scattering. The adjusted excitation energies are shown in
Fig.~\ref{Figdispersion}a, illustrating that magnetic spectral
weight generally shifts to lower energies in an applied field.
However, due to the $H_{\rm st}$-induced gap, this effect is much
less pronounced than in a uniform field as the ground state energy
increases.\cite{Muller}\par

\section{Mean-field theory}

We now present a simple mean-field theory of interacting
$S$=$\frac{1}{2}$ fermions which captures both the emergence of a
new excitation at high energies and the dispersion of the bound
spinon states. As it is well known, a mean-field approach usually is
not adequate to solve a one dimensional system. This is because, for
finite range interactions, the fluctuations are strong enough to
preclude a non-zero value of an order parameter at any fine
temperature. Moreover, if the order parameter is associated with a
continuous symmetry, its mean value is zero even at T=0.
Consequently, a mean field theory that assumes a non-zero value of
the order parameter with excitations that are originated by small
fluctuations of such quantity cannot be a good description of a
general one dimensional system.

In the present case, the system is invariant under rotations around
the $z$ axis for $H_{\rm st}=0$ and the candidate to be the order
parameter is the $xy$ planar component of the staggered
magnetization $M^{\rm st}_{\perp}$. As expected for a continuous
symmetry and a gapless spectrum, the system is critical at $T=0$,
i.e., the order parameter has divergent fluctuations. However, the
staggered field $H_{\rm st}=0$ couples linearly with the order
parameter $M^{\rm st}_{x}$ along the $x$ direction. Consequently,
for $H_{\rm st}\neq0$, the U(1) rotational symmetry is explicitly
broken and the mean value of $M^{\rm st}_{x}$ becomes non-zero at
any temperature. This also changes the nature of the excitations.
The spinons (kink and antikinks) are no longer the low energy
quasiparticles of the system since a pair of them is now confined by
a linear potential. A similar effect occurs when we increase the
dimension of the system due to the interchain interaction.
Therefore, we expect a mean-field treatment to be good approach for
high enough values of $H_{\rm st}$.

For $H_{\rm st}=0$, we know that the theory must be critical with
the associated linear soft modes shown in Fig.\ref{Figdispersion}a.
We also know that one dimensional spin system can be mapped into a
fermionic system. In addition, a non-interacting fermionic
Hamiltonian has a ground state that is also critical and has linear
soft modes like the ones shown in Fig.\ref{Figdispersion}a.
Therefore, it is convenient to use a fermionic representation for
the mean-field approach. For this purpose, the spin degrees of ${\rm
Cu^{2+}}$ ions are described in terms of fermionic creation and
annihilation operators:
\begin{equation}
S^{\nu}_j=\frac{1}{2}\sum_{\alpha,\alpha'}
c^{\dagger}_{j\alpha} \sigma^{\nu}_{\alpha \alpha'} c^{\;}_{j\alpha'}\, ,
\end{equation}
where $\nu={x,y,z}$ and $\sigma^{\nu}$ are the Pauli matrices. Using
this fermionic representation, Baskaran, Zou, and Anderson
\cite{Baskaran} proposed a mean-field theory (MFT) to treat low
dimensional Heisenberg spin $S=1/2$ Hamiltonians. The MFT was
generalized to SU(N) spin models (for large $N$) by Affleck and
Marston.\cite{Affleck_Marston} Arovas and Auerbach \cite{Arovas}
studied this theory in comparison with the Bethe Ansatz solution and
showed that the fluctuation corrections are important in enforcing
the Gutzwiller projection. Using an extended version of this MFT, we
will study here the ground state properties and the spin dynamics of
the $S$=$\frac{1}{2}$ chain in a staggered field given by
$\mathcal{H}$.\par

Apart form an irrelevant constant, the expression for $\mathcal{H}$
in the fermionic representation is:
\begin{equation}
    \mathcal{H}=- \frac{J}{2} \sum_{i,\sigma,\sigma'}
    c^{\dagger}_{i\sigma} c{\;}_{i+1 \sigma} c^{\dagger}_{i+1 \sigma'}c{\;}_{i\sigma'}
    -\frac{g_c}{2} \mu_B H \sum_{i,\sigma} \sigma n_{i\sigma} -
    \frac{g_a}{2} \mu_B H_{\rm st} \sum_{i,\sigma} (-1)^i c^{\dagger}_{i\sigma} c{\;}_{i {\bar \sigma}}\, ,
\end{equation}with ${\bar \sigma}= -\sigma$. Since there is one
spin per site, there is a constraint on the fermion occupation
number: $n_i=\sum_{\sigma}n_{i\sigma}=1$ with
$n_{i\sigma}=c^{\dagger}_{i\sigma}c^{\;}_{i\sigma}$. For the
Heisenberg term, we will use a linear combination of the mean-field
decoupling introduced in Ref.~\onlinecite{Baskaran} and the other
natural decoupling in the presence of a staggered field along the
$x$ direction:
\begin{equation}
\mathcal{H}_{MF}=-\frac{J \gamma}{2}\sum_{i\sigma}
(c^{\dagger}_{i\sigma} c^{\;}_{i+1 \sigma}+ {\rm H.c.}) -
\frac{g_c}{2} \mu_B H
    \sum_{i,\sigma} \sigma n_{i\sigma} - \frac{1}{2} (g_a \mu_B H_{\rm st} + J \delta)
    \sum_{i,\sigma} (-1)^i c^{\dagger}_{i\sigma} c{\;}_{i {\bar \sigma}}+ \lambda \sum_{i} n_i\ ,
\end{equation}
where the Lagrange multiplier or chemical potential, $\lambda$,
enforces the constraint of one spin per site at mean-field level. We
are assuming translational invariance for $\gamma_{i}$,
$\gamma=\sum_{\sigma} \langle
c^{\dagger}_{i\sigma}c_{i+1\sigma}\rangle=\gamma_i$, and a staggered
dependence for the effective field $\delta_i=\sum_{\sigma} \langle
c^{\dagger}_{i\sigma}c_{i{\bar \sigma}}\rangle=\delta (-1)^i$. This
staggered dependence is induced by the field $H_{\rm st}$. In
momentum space, this leads to
\begin{equation}
    \mathcal{H}_{MF}=\sum_{-\pi< k \leq \pi, \sigma}\left[ \left(-J \gamma \cos(k) -
    \frac{\sigma}{2} g_c \mu_B H\right) c^+_{k\sigma}c_{k\sigma}
    - \frac{1}{2} (g_a \mu_B H_{\rm st} + J \delta) (c^+_{k+\pi \sigma}c_{k{\bar \sigma}}
     +c^+_{k{\bar \sigma}}c_{k+\pi \sigma})\right]\, ,
\end{equation}which can be written in the matrix formulation as
\begin{equation}
    H_{MF}(k)_\sigma = \left[
    \begin{array}{cc}-J\gamma \cos(k) - \sigma \frac{1}{2} g_c \mu_B H &- \frac{1}{2}
    (g_a \mu_B H_{\rm st} + J \delta)\\
    - \frac{1}{2} (g_a \mu_B H_{\rm st} + J \delta)&J\gamma\cos(k)+\sigma \frac{1}{2} g_c \mu_B
    H\end{array}
\right]\, ,
\end{equation} for $-\pi/2 < k \leq \pi/2$. The eigenvalues of this matrix,
\begin{equation}
    \epsilon^{\pm}_{k \sigma} = \pm \sqrt{\left(J\gamma\cos(k)+\sigma \frac{g_c}{2}
    \mu_B H \right)^2 + \frac{1}{4}(g_a \mu_B H_{\rm st}+J \delta)^2 }\, ,
    \label{Eqfermiondispersion}
\end{equation}
are the energies of the quasi-particle operators,
\begin{eqnarray}
\alpha^{\dagger}_{k\sigma} = u_{k \sigma} c^{\dagger}_{k\sigma}+
v_{k \sigma} c^{\dagger}_{k+\pi\bar \sigma}
\nonumber  \\
\beta^{\dagger}_{k \sigma} = - v_{k \sigma} c^{\dagger}_{k\sigma}+
u_{k \sigma} c^{\dagger}_{k+\pi \bar \sigma}\, , \label{qpo1}
\end{eqnarray}
with
\begin{eqnarray}
u_{k\sigma} = \frac{\epsilon^{+}_{k \sigma}+J\gamma\cos(k)+\sigma
\frac{1}{2} g_c
    \mu_B H}{\sqrt{\left(\epsilon^{+}_{k \sigma}+J\gamma\cos(k)+\sigma
    \frac{1}{2}g_c
    \mu_B H \right)^2+ \frac{1}{4} (g_a \mu_B H_{\rm st}+J \delta)^2}}\, ,
\nonumber  \\
v_{k\sigma} = \frac{\frac{1}{2} g_a \mu_B H_{\rm st}+ J
\delta}{\sqrt{\left(\epsilon^{+}_{k \sigma}+J\gamma\cos(k)+\sigma
\frac{1}{2} g_c \mu_B H \right)^2+ \frac{1}{4} (g_a \mu_B H_{\rm
st}+J \delta)^2 }}\, . \label{qpo2}
\end{eqnarray}
$H_{MF}$ is diagonal in the new basis:
\begin{equation}
\mathcal{H}_{MF}=\sum_{-\pi/2 < k \leq \pi/2, \sigma}
(\epsilon^+_{k\sigma} \beta^{\dagger}_{k\sigma}\beta^{\;}_{k\sigma}
+\epsilon^-_{k\sigma}
\alpha^{\dagger}_{k\sigma}\alpha^{\;}_{k\sigma})\, .
\end{equation}
The mean-field  parameters $\gamma$ and $\delta$ are given by the
self-consistent equations:
\begin{eqnarray}
\gamma=\frac {1}{2\pi} \sum_{\sigma} \int_{-\pi/2}^{\pi/2} \cos(k)
(u_{k\sigma}^2-v_{k \sigma}^2) [\langle  \alpha^{\dagger}_{k\sigma}
\alpha^{\;}_{k\sigma} \rangle - \langle  \beta^{\dagger}_{k\sigma}
\beta^{\;}_{k\sigma} \rangle] dk\, ,
\nonumber \\
\delta=\frac {1}{2\pi} \sum_{\sigma} \int_{-\pi/2}^{\pi/2}
u_{k\sigma} v_{k \sigma}
[\langle  \alpha^{\dagger}_{k\sigma} \alpha^{\;}_{k\sigma} \rangle
- \langle  \beta^{\dagger}_{k\sigma} \beta^{\;}_{k\sigma} \rangle] dk \, .
\label{self}
\end{eqnarray}
The value of $\lambda$ is determined by imposing the average
occupation per site to be equal to 1:
\begin{equation}
\frac {1}{2\pi} \sum_{\sigma} \int_{-\pi/2}^{\pi/2}
[\langle  \alpha^{\dagger}_{k\sigma} \alpha^{\;}_{k\sigma} \rangle
+ \langle  \beta^{\dagger}_{k\sigma} \beta^{\;}_{k\sigma} \rangle] dk =1 \,.
\end{equation}
When $H=H_{\rm st}=0$, the integration over the phase fluctuations
of the local field $\gamma_i$ \cite{Arovas} renormalizes the value
of $\gamma$ given by Eq.(\ref{self}): ${\tilde \gamma} = \pi
\gamma/\sqrt{2} $. This renormalization improves considerably the
comparison with the exact des-Cloizeaux-Pearson
\cite{Cloizeaux_Pearson} two-spinon threshold ($\gamma_{ex}=\pi/2$).
To improve the quantitative comparison of the MFT with the
experiment and the exact diagonalization results, we will assume
here that the same renormalization factor, $\pi / \sqrt{2}$, must be
applied when $H$ and $H_{\rm st}$ are finite. The value of $\gamma$
is a measure of the effective strength of the spin fluctuations
introduced by the Heisenberg term. More specifically, $\gamma J$ is
the Fermi velocity that in the original spin language corresponds to
the velocity of the spinon excitations.

\begin{figure}[ht]
\begin{center}
  \includegraphics[height=9cm,bbllx=220,bblly=280,bburx=575,
  bbury=710,angle=0,clip=]{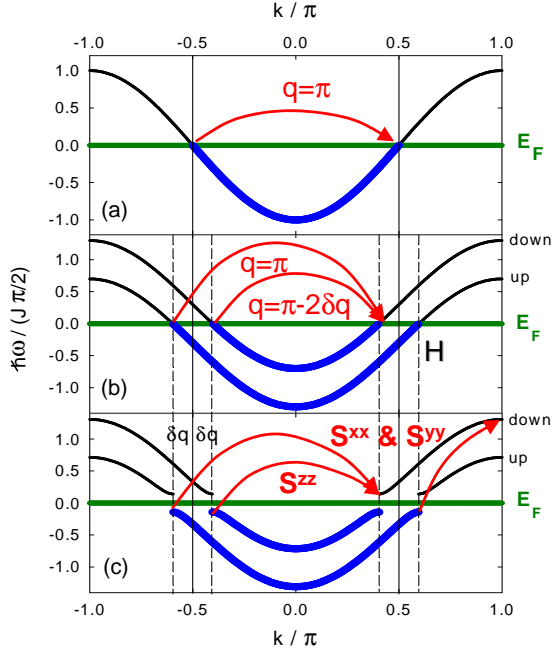}
  \vspace{0.2cm}
  \caption{Schematic representation of the Fermi particle dispersion
  in (a) zero field, (b) uniform field $H$ and (c)
  uniform field $H$ and staggered field $H_{\rm st}$ with $H > H_{\rm st}$.
  In zero field the spin up and down particles with wave-vector $k$ have
  the same dispersion.}
  \label{FigFermi}
\end{center}
\end{figure}

Fig.~\ref{FigFermi} shows the evolution of the fermionic bands when
the uniform and the staggered fields are applied. In the absence of
these magnetic fields (Fig.~\ref{FigFermi}a), there is only one band
and the two-spinon cross section is associated with the continuum of
particle-hole excitations. The dispersion relation for the lower
branch is ${\tilde \gamma} J |\sin (q)|$. When a uniform magnetic
field $H\neq0$ is applied (Fig.~\ref{FigFermi}b), the spin up and
down bands are split by the Zeeman term. As a consequence, there is
a change $\delta q$ of Fermi wave vectors $|q_F|=\pi/2\pm \delta q$
and a corresponding change in the wave vectors of the zero energy
modes:  the energy of the  transverse modes goes to zero  at $q=\pi$
and $q=2\delta q$, while the longitudinal excitations have gapless
modes at $q=0$ and  $q=\pi - 2\delta q$. From
Eq.~(\ref{Eqfermiondispersion}), the main effect of a non-zero
staggered field $H_{\rm st}$ is to open a gap at the Fermi level,
i.e., the fermionic system becomes an insulator and the spectrum is
gaped for any excitation (Fig.~\ref{FigFermi}c). The gap results
from the inter-band scattering which is introduced by the staggered
field $H_{\rm st}$. According to Eqs.~(\ref{qpo1}), the degree of
mixing is maximum at $q=q_F$. The emergence of this gap is
consistent with the experimental data shown in
Fig.~\ref{Figdispersion}.

To study the excitations of the new ground state induced by $H_{\rm
st}$ it is necessary to calculate the neutron scattering cross
section within our MFT. The neutron scattering cross section for the
transverse excitations ($\nu=x,y$) at $T=0\;\mathrm{K}$ is given by:
\begin{eqnarray}
S^{\nu \nu}(q,\omega)=\frac{1}{8\pi} \sum_{\sigma}
[ \int^{\frac{\pi}{2}}_{\frac{\pi}{2}-q}
(u_{k+q \sigma}u_{k \sigma}+ \eta v_{k+q\sigma}v_{k\sigma})^2
\delta(\omega-\epsilon^{+}_{k+q\sigma}+\epsilon^{-}_{k \sigma}) dk
\nonumber \\
+ \int^{\frac{\pi}{2}-q}_{-\frac{\pi}{2}}
(u_{k+q \sigma}v_{k {\bar \sigma}}+ \eta v_{k+q \sigma}u_{k{\bar \sigma}})^2
\delta(\omega-\epsilon^{+}_{k+q\sigma}+\epsilon^{-}_{{k \bar \sigma}})] dk\, ,
\label{cross}
\end{eqnarray}
where $\eta=-1$ for $\nu=x$, $\eta=1$ for $\nu=y$ and $-\pi < q \leq \pi$. Note that
in Eq.(\ref{cross}), $k+q$ must be contracted to the reduced Brillouin zone
($k+q \equiv k+q+n\pi$). The cross section for longitudinal excitations is:
\begin{eqnarray}
S^{zz}(q,\omega)=\frac{1}{8\pi} \sum_{\sigma}
[ \int^{\frac{\pi}{2}}_{\frac{\pi}{2}-q}
(u_{k+q {\bar \sigma}}u_{k\sigma} + v_{k+q{\bar \sigma}}v_{k\sigma})^2
\delta(\omega-\epsilon^{+}_{k+q {\bar \sigma}}+\epsilon^{-}_{k \sigma})
\nonumber \\
+ \int^{\frac{\pi}{2}-q}_{-\frac{\pi}{2}} (u_{k+q \sigma} v_{k
\sigma} + v_{k+q \sigma}u_{k\sigma})^2
\delta(\omega-\epsilon^{+}_{k+q \sigma}+\epsilon^{-}_{k \sigma})]
dk\, . \label{long}
\end{eqnarray}

\begin{figure}[ht]
\begin{center}
\includegraphics[angle=-90,width=7.0cm,scale=1.2]{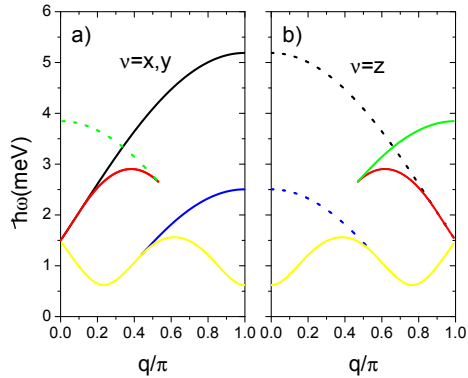}
\vspace{-0.2cm} \caption{ a) Transverse  and b) longitudinal
excitations obtained from the mean-field theory, as described in the
text in more detail.} \label{mfbr}
\end{center}
\end{figure}

Equations (\ref{cross}) and  (\ref{long}) reveal well-defined
excitations that are determined by the cancelation of $d\omega/dk$,
i.e., the divergence of the Jacobian. In Fig.~\ref{mfbr}a, we show
the different branches of the transverse excitations. The lower
branch corresponds to the dispersion relation of the low energy
excitations. In agreement with the experiment (see
Fig.~\ref{Figdispersion}), there are two minima, one is located at
the incommensurate wave vector $q_I=2\arcsin(g_c \mu_B H/2\gamma J)$
and the other one occurs at $q=\pi$. It is interesting to note that
for $\nu=x$ the intensity of this branch goes to zero at $q_I=2
\delta q$ due to a cancelation of the matrix element that multiplies
the delta function in the integrand of Eq.~(\ref{cross}). The black
and the blue curves of Fig.~\ref{mfbr}a are the upper boundaries of
inter-band particle-hole excitations associated with the transverse
modes (Fig.~\ref{FigFermi}b). More specifically, the black curve
results from excitations in which an electron is annihilated in the
lower band and created in the upper band, while for the blue curve
the process is the opposite. The green curve of Fig.~\ref{mfbr}b is
the upper boundary for the intra-band particle-hole excitations that
describes the longitudinal modes. Note that these boundaries appear
with dashed lines in the spectrum of excitations with the other
polarization (Fig.~\ref{FigFermi}). This is a consequence of the
inter-band $q=\pi$  scattering which is introduced by the staggered
field $H_{\rm st}$. The dashed line just indicates that these
``shadow'' branches have a very small intensity.\par

The gap in the low energy spectrum is not the only qualitative
change introduced by the staggered field $H_{\rm st}$ within the
mean-field approach. Since the Fermi wave vectors  $q_F=\pm \pi/2\pm
\delta q$ are now extremal points of the new bands (see
Fig.~\ref{FigFermi}c), we expect the emergence of a new branch of
transverse excitations associated with transitions between points
that are close to $q_F=\mp \pi/2\pm \delta q$ ($q_F=\pm \pi/2\pm
\delta q$) and points in the proximity of  $q=0$ ($q=\pi$). This new
branch of excitations (see the red curve in Fig.~\ref{mfbr}a)
coincides with the new excitation which is experimentally observed
at high energies (see Fig.~\ref{Fignewexcitation}), and thus
explains our experimental results. For the longitudinal
polarization, the new branch of excitations is shifted by $\pi$
relative to the transverse polarization (see the red curve in
Fig.~\ref{mfbr}b). Since the maximum of this longitudinal branch is
located at $q=\pi/2+\delta q$, it is difficult to distinguish this
branch from the breathers in the experimental data. The energy of
this maximum is close to $3 \mathrm{meV}$ according to the MFT while
the experimental value is $3.4 \mathrm{meV}$. This is reasonable
given that the MFT is not an adequate approximation to give a
quantitative description of the excitations.\par

\section{Exact diagonalization of finite length chains}

The intensities of the different branches are not properly described
by the mean-field equations (\ref{cross}) and (\ref{long}). For
instance, the MFT predicts a high intensity for the upper boundary
of the two-spinon excitations even at zero field for which very
accurate calculations are available.\cite{Muller} This is clearly an
artifact of the MFT. To obtain an accurate description of the
intensities and the energies of the different branches we
complemented our analytical approach with the exact diagonalization
of finite size chains. Using the Lanczos method, we obtained the
exact ground state of $H$ for finite chains of length
$L=12,14,16,18,22,20,24$. Having the ground state, we  computed the
dynamical magnetic susceptibility, $\chi(\omega,q)$, for all the
possible wave vectors $q=0,\, 2\pi/L,\,....,\,2\pi (L-1)/L$ of a
chain of length $L$ using the method introduced in
Ref.~\onlinecite{Gagliano_Balseiro}. The wave vector $q=\pi$ is
present in all of the considered chains. The small changes in the
calculated $\chi(\omega,\pi)$ as a function of $L$ indicate that the
finite size effects are small for the considered problem. In
general, smaller finite size effects are expected for systems that
have an excitation gap because the spin-spin correlation length is
finite.\par

The $T=0\;\mathrm{K}$ structure factors were calculated using
\begin{equation}
    S^{\alpha\alpha}(q,\omega)=\frac{1}{\pi} \chi''^{\alpha\alpha}(q,\omega)\,
    .
\end{equation}The energy spectra were obtained
by convoluting the discrete spectra for finite chains with
Lorentzian functions with a full width at half maximum, $2\Gamma=0.1
\mathrm{meV}$, in order to model the experimental energy resolution.
The intensity of the calculated structure factors is given for a
chain of $L$ spins and normalized so that $\sum_{q,\alpha} \int
d\omega S^{\alpha\alpha}(q,\omega)= S(S+1)$ as required by the total
scattering sum rule.\par

\begin{figure}[ht]
\begin{center}
  \includegraphics[height=7cm,bbllx=27,bblly=251,bburx=575,
  bbury=564,angle=0,clip=]{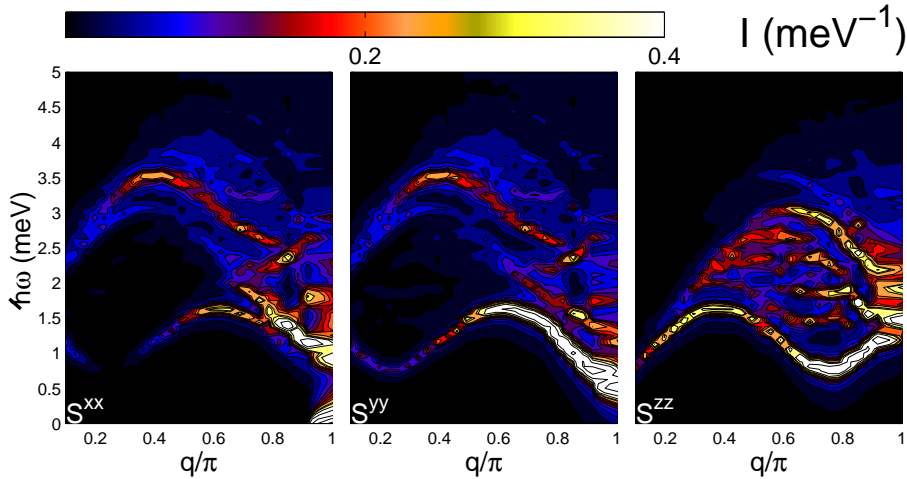}
  \caption{The dynamic structure factors $S^{\alpha \alpha}(q,\omega)$
  obtained through exact diagonalization of finite chains for
  $H_{\rm st}$=$0.075H$, where $\alpha$=$x$, $y$ and $z$ are the spin
  polarizations. The uniform field is along the c-axis (z polarization),
  the staggered field along the a-axis (x polarization).}
  \label{Figexactdiag}
\end{center}
\end{figure}

The calculated neutron scattering for all chain lengths was averaged
and the dynamic structure factor for the three different
polarizations is shown in Fig.~\ref{Figexactdiag} as a function of
wave-vector transfer and energy transfer for $H_{\rm st}$=$0.075H$
on an absolute scale. The structure factor $S^{zz}$ polarized along
the uniform field contains a well-defined excitation with a minimum
gap energy at an incommensurate wave-vector. The structure factors
$S^{xx}$ and $S^{yy}$ polarized perpendicular to the uniform field
contain well-defined excitations whose dispersion has a minimum at
the antiferromagnetic point, with $S^{yy}$ having an excitation at
lower energy than $S^{xx}$. These excitations correspond to the
first and second breather of the quantum sine-Gordon model.\par

Fig.~\ref{Figexactdiag} also provides evidence that the excitation
spectrum contains a substantial amount of continuum states, as
observed in the experiment.\cite{Kenzelmann_CDC_PRL} These states
lie higher in energy than the well-defined low-energy excitations
and extend to high energies. At $H_{\rm st}$=$0.075H$ our numerical
calculations yield an expectation value of the Hamiltonian per
$S=\frac{1}{2}$ of $\mathcal{H}=-0.34 J$, increased from the ground
state energy at zero field, $\mathcal{H}=(\frac{1}{4} - \ln 2 ) J=
-0.44 J$, by $0.1 J$.\par

\begin{figure}[ht]
\begin{center}
  \includegraphics[height=7cm,bbllx=24,bblly=254,bburx=550,
  bbury=548,angle=0,clip=]{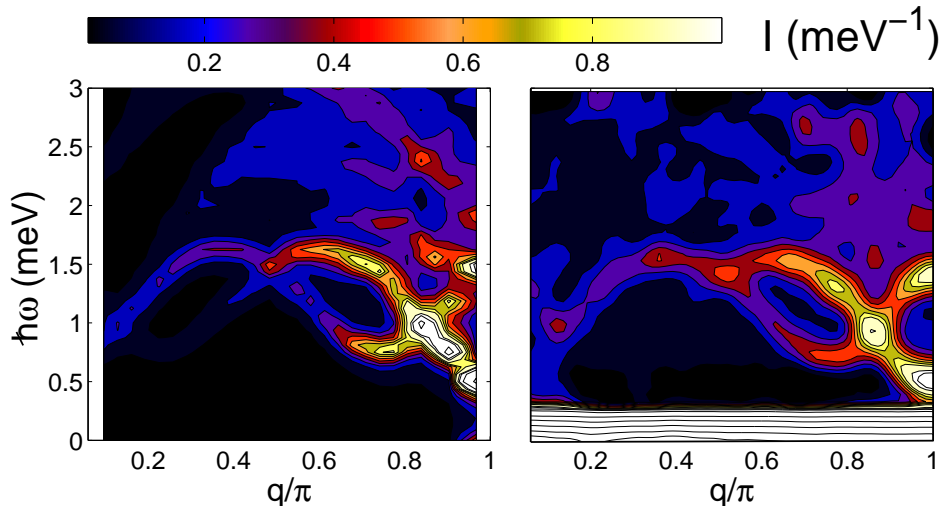}
  \caption{(a) Calculated dynamic structure factor $S(q,\omega)$
  for wave-vectors and polarization of the DCS experiment.
  (b) Structure factor measured using DCS
  (Ref.~\protect\onlinecite{Kenzelmann_CDC_PRL}).}
  \label{Figcomparison}
\end{center}
\end{figure}

The numerical data were binned and Gaussian fits were used to obtain
excitation energies as a function of wave-vector. These results are
shown in Fig.~\ref{Figdispersion}a as a solid line, showing that the
numerical calculations reproduce the dispersion relation of the low
energy modes. We calculated the dynamic structure factor for the DCS
experiment taking into account wave-vector dependent mixing of the
polarized dynamic structure factor $S^{\alpha\alpha}$.
Fig.~\ref{Figcomparison} directly compares the calculated and
measured intensities on an absolute scale, showing that there is
excellent agreement between the numerical calculations and the
experiment.\par

In addition, Figure~\ref{Figexactdiag} shows that a new branch of
transverse magnetic excitations with the maximum around $q=0.4\pi$
and $\hbar \omega=3.5\mathrm{meV}$ emerges when the staggered field
is present. This provides a quantitative explanation for the high
energy peak that appears in the neutron scattering data (see
Fig.~\ref{Fignewexcitation}). Muller {\it et al}.~\cite{Muller}
showed that this branch is not present for $H_{\rm st}=0$ by using
macroscopic selection rules. For non-zero $H_{\rm st}$, the total
spin $S$ and its projection along the $z$-direction, $S^{z}$, are
not good quantum numbers anymore, allowing the emergence of this new
excitation. Our mean-field approach is in good agreement with this
result.\par

\section{Conclusions}

In summary, we have performed neutron scattering experiments,
numerical calculations and an analytical study investigating the AF
$S$=$\frac{1}{2}$ chain in uniform and staggered fields. We found
that the incommensurate bound-spinon states are well described by a
mapping to an interacting fermionic model after a renormalization of
the energy. The model also explains the emergence of a new
excitations with the application of a staggered field. Our results
suggest that the proposed mapping is more powerful than initial
results suggested, and that it may also be useful for other quantum
spin systems with a relatively short correlation length.\par

\begin{acknowledgments}
Work at JHU was supported by the NSF through DMR-0306940. DCS and
the high-field magnet at NIST were supported in part by the NSF
through DMR-0086210 and DMR-9704257. C. B. and D.~H. R. gratefully
acknowledge discussions with A.~J. Millis concerning the mean-field
theory of $S$=$\frac{1}{2}$ chains.
\end{acknowledgments}



\begin{thebibliography}{22}
\expandafter\ifx\csname
natexlab\endcsname\relax\def\natexlab#1{#1}\fi
\expandafter\ifx\csname bibnamefont\endcsname\relax
  \def\bibnamefont#1{#1}\fi
\expandafter\ifx\csname bibfnamefont\endcsname\relax
  \def\bibfnamefont#1{#1}\fi
\expandafter\ifx\csname citenamefont\endcsname\relax
  \def\citenamefont#1{#1}\fi
\expandafter\ifx\csname url\endcsname\relax
  \def\url#1{\texttt{#1}}\fi
\expandafter\ifx\csname urlprefix\endcsname\relax\def\urlprefix{URL
}

\bibitem[{\citenamefont{Faddeev and Takhtajan}(1981)}]{Faddeev_Takhtajan}
\bibinfo{author}{\bibfnamefont{L.~D.} \bibnamefont{Faddeev}} \bibnamefont{and}
  \bibinfo{author}{\bibfnamefont{L.~A.} \bibnamefont{Takhtajan}},
  \bibinfo{journal}{Phys. Lett. A} \textbf{\bibinfo{volume}{85}},
  \bibinfo{pages}{375} (\bibinfo{year}{1981}).

\bibitem[{\citenamefont{Tennant
  et~al.}(1995{\natexlab{a}})\citenamefont{Tennant, Cowley, Nagler, and
  Tsvelik}}]{Tennant95}
\bibinfo{author}{\bibfnamefont{D.~A.} \bibnamefont{Tennant}},
  \bibinfo{author}{\bibfnamefont{R.~A.} \bibnamefont{Cowley}},
  \bibinfo{author}{\bibfnamefont{S.~E.} \bibnamefont{Nagler}},
  \bibnamefont{and} \bibinfo{author}{\bibfnamefont{A.~M.}
  \bibnamefont{Tsvelik}}, \bibinfo{journal}{Phys. Rev. B}
  \textbf{\bibinfo{volume}{52}}, \bibinfo{pages}{13368}
  (\bibinfo{year}{1995}{\natexlab{a}}).

\bibitem[{\citenamefont{Kenzelmann et~al.}(2001)\citenamefont{Kenzelmann,
  Zheludev, Raymond, Ressouche, Masuda, B\"{o}ni, Kakurai, Tsukada, Uchinokura,
  and Coldea}}]{Kenzelmann_BaCu2Si2O7}
\bibinfo{author}{\bibfnamefont{M.}~\bibnamefont{Kenzelmann}},
  \bibinfo{author}{\bibfnamefont{A.}~\bibnamefont{Zheludev}},
  \bibinfo{author}{\bibfnamefont{S.}~\bibnamefont{Raymond}},
  \bibinfo{author}{\bibfnamefont{E.}~\bibnamefont{Ressouche}},
  \bibinfo{author}{\bibfnamefont{T.}~\bibnamefont{Masuda}},
  \bibinfo{author}{\bibfnamefont{P.}~\bibnamefont{B\"{o}ni}},
  \bibinfo{author}{\bibfnamefont{K.}~\bibnamefont{Kakurai}},
  \bibinfo{author}{\bibfnamefont{I.}~\bibnamefont{Tsukada}},
  \bibinfo{author}{\bibfnamefont{K.}~\bibnamefont{Uchinokura}},
  \bibnamefont{and} \bibinfo{author}{\bibfnamefont{R.}~\bibnamefont{Coldea}},
  \bibinfo{journal}{Phys. Rev. B} \textbf{\bibinfo{volume}{64}},
  \bibinfo{pages}{054422} (\bibinfo{year}{2001}).

\bibitem[{\citenamefont{Stone et~al.}(2003)\citenamefont{Stone, Reich, Broholm,
  Lefmann, Rischel, Landee, and Turnbull}}]{Stone}
\bibinfo{author}{\bibfnamefont{M.~B.} \bibnamefont{Stone}},
  \bibinfo{author}{\bibfnamefont{D.~H.} \bibnamefont{Reich}},
  \bibinfo{author}{\bibfnamefont{C.}~\bibnamefont{Broholm}},
  \bibinfo{author}{\bibfnamefont{K.}~\bibnamefont{Lefmann}},
  \bibinfo{author}{\bibfnamefont{C.}~\bibnamefont{Rischel}},
  \bibinfo{author}{\bibfnamefont{C.~P.} \bibnamefont{Landee}},
  \bibnamefont{and} \bibinfo{author}{\bibfnamefont{M.~M.}
  \bibnamefont{Turnbull}}, \bibinfo{journal}{Phys. Rev. Lett.}
  \textbf{\bibinfo{volume}{91}}, \bibinfo{pages}{037205}
  (\bibinfo{year}{2003}).

\bibitem[{\citenamefont{Schulz}(1996)}]{Schulz96}
\bibinfo{author}{\bibfnamefont{H.~J.} \bibnamefont{Schulz}},
  \bibinfo{journal}{Phys. Rev. Lett.} \textbf{\bibinfo{volume}{77}},
  \bibinfo{pages}{2790} (\bibinfo{year}{1996}).

\bibitem[{\citenamefont{Tennant
  et~al.}(1995{\natexlab{b}})\citenamefont{Tennant, , Nagler, Welz, Shirane,
  and Yamada}}]{Tennant95/2}
\bibinfo{author}{\bibfnamefont{D.~A.} \bibnamefont{Tennant}}, ,
  \bibinfo{author}{\bibfnamefont{S.~E.} \bibnamefont{Nagler}},
  \bibinfo{author}{\bibfnamefont{D.}~\bibnamefont{Welz}},
  \bibinfo{author}{\bibfnamefont{G.}~\bibnamefont{Shirane}}, \bibnamefont{and}
  \bibinfo{author}{\bibfnamefont{K.}~\bibnamefont{Yamada}},
  \bibinfo{journal}{Phys. Rev. B} \textbf{\bibinfo{volume}{52}},
  \bibinfo{pages}{13381} (\bibinfo{year}{1995}{\natexlab{b}}).

\bibitem[{\citenamefont{Zheludev et~al.}(2000)\citenamefont{Zheludev,
  Kenzelmann, Raymond, Ressouche, Masuda, Kakurai, Maslov, Tsukada, Uchinokura,
  and Wildes}}]{Zheludev_BaCu2Si2O7}
\bibinfo{author}{\bibfnamefont{A.}~\bibnamefont{Zheludev}},
  \bibinfo{author}{\bibfnamefont{M.}~\bibnamefont{Kenzelmann}},
  \bibinfo{author}{\bibfnamefont{S.}~\bibnamefont{Raymond}},
  \bibinfo{author}{\bibfnamefont{E.}~\bibnamefont{Ressouche}},
  \bibinfo{author}{\bibfnamefont{T.}~\bibnamefont{Masuda}},
  \bibinfo{author}{\bibfnamefont{K.}~\bibnamefont{Kakurai}},
  \bibinfo{author}{\bibfnamefont{S.}~\bibnamefont{Maslov}},
  \bibinfo{author}{\bibfnamefont{I.}~\bibnamefont{Tsukada}},
  \bibinfo{author}{\bibfnamefont{K.}~\bibnamefont{Uchinokura}},
  \bibnamefont{and} \bibinfo{author}{\bibfnamefont{A.}~\bibnamefont{Wildes}},
  \bibinfo{journal}{Phys. Rev. Lett.} \textbf{\bibinfo{volume}{85}},
  \bibinfo{pages}{4799} (\bibinfo{year}{2000}).

\bibitem[{\citenamefont{Kenzelmann et~al.}(2004)\citenamefont{Kenzelmann, Chen,
  Broholm, Reich, and Qiu}}]{Kenzelmann_CDC_PRL}
\bibinfo{author}{\bibfnamefont{M.}~\bibnamefont{Kenzelmann}},
  \bibinfo{author}{\bibfnamefont{Y.}~\bibnamefont{Chen}},
  \bibinfo{author}{\bibfnamefont{C.}~\bibnamefont{Broholm}},
  \bibinfo{author}{\bibfnamefont{D.~H.} \bibnamefont{Reich}}, \bibnamefont{and}
  \bibinfo{author}{\bibfnamefont{Y.}~\bibnamefont{Qiu}},
  \bibinfo{journal}{Phys. Rev. Lett.} \textbf{\bibinfo{volume}{93}},
  \bibinfo{pages}{017204} (\bibinfo{year}{2004}).

\bibitem[{\citenamefont{Affleck and Oshikawa}(1999)}]{Affleck_Oshikawa}
\bibinfo{author}{\bibfnamefont{I.}~\bibnamefont{Affleck}} \bibnamefont{and}
  \bibinfo{author}{\bibfnamefont{M.}~\bibnamefont{Oshikawa}},
  \bibinfo{journal}{Phys. Rev. B} \textbf{\bibinfo{volume}{60}},
  \bibinfo{pages}{1038} (\bibinfo{year}{1999}).

\bibitem[{\citenamefont{Chen and \textit{et al.}}()}]{Chen_CDC}
\bibinfo{author}{\bibfnamefont{Y.}~\bibnamefont{Chen}} \bibnamefont{and}
  \bibinfo{author}{\bibnamefont{\textit{et al.}}}, \bibinfo{howpublished}{to be
  published}.

\bibitem[{\citenamefont{Landee et~al.}(1987)\citenamefont{Landee, Lamas,
  Greeney, and B\"{u}cher}}]{Landee}
\bibinfo{author}{\bibfnamefont{C.~P.} \bibnamefont{Landee}},
  \bibinfo{author}{\bibfnamefont{A.~C.} \bibnamefont{Lamas}},
  \bibinfo{author}{\bibfnamefont{R.~E.} \bibnamefont{Greeney}},
  \bibnamefont{and} \bibinfo{author}{\bibfnamefont{K.~G.}
  \bibnamefont{B\"{u}cher}}, \bibinfo{journal}{Phys. Rev. B}
  \textbf{\bibinfo{volume}{35}}, \bibinfo{pages}{228} (\bibinfo{year}{1987}).

\bibitem[{\citenamefont{Willett and Chang}(1970)}]{Willett_Chang}
\bibinfo{author}{\bibfnamefont{R.~D.} \bibnamefont{Willett}} \bibnamefont{and}
  \bibinfo{author}{\bibfnamefont{K.}~\bibnamefont{Chang}},
  \bibinfo{journal}{Inorg. Chem. Acta} \textbf{\bibinfo{volume}{4}},
  \bibinfo{pages}{447} (\bibinfo{year}{1970}).

\bibitem[{\citenamefont{Sato and Oshikawa}(2004)}]{Oshikawanew}
\bibinfo{author}{\bibfnamefont{M.} \bibnamefont{Sato}} \bibnamefont{and}
  \bibinfo{author}{\bibfnamefont{M.}~\bibnamefont{Oshikawa}},
  \bibinfo{journal}{Phys. Rev. B} \textbf{\bibinfo{volume}{69}},
  \bibinfo{pages}{054406} (\bibinfo{year}{2004}).

\bibitem[{\citenamefont{Cooper and Nathans}(1967)}]{Cooper_Nathans}
\bibinfo{author}{\bibfnamefont{M.~J.} \bibnamefont{Cooper}} \bibnamefont{and}
  \bibinfo{author}{\bibfnamefont{R.}~\bibnamefont{Nathans}},
  \bibinfo{journal}{Acta Crys.} \textbf{\bibinfo{volume}{23}},
  \bibinfo{pages}{357} (\bibinfo{year}{1967}).

\bibitem[{\citenamefont{Bougourzi et~al.}(1998)\citenamefont{Bougourzi,
  Karbach, and M\"{u}ller}}]{Bougourzi_Karbach}
\bibinfo{author}{\bibfnamefont{A.~H.} \bibnamefont{Bougourzi}},
  \bibinfo{author}{\bibfnamefont{M.}~\bibnamefont{Karbach}}, \bibnamefont{and}
  \bibinfo{author}{\bibfnamefont{G.}~\bibnamefont{M\"{u}ller}},
  \bibinfo{journal}{Phys. Rev. B} \textbf{\bibinfo{volume}{57}},
  \bibinfo{pages}{11429} (\bibinfo{year}{1998}).

\bibitem[{\citenamefont{M\"{u}ller et~al.}(1981)\citenamefont{M\"{u}ller,
  Thomas, Beck, and Bonner}}]{Muller}
\bibinfo{author}{\bibfnamefont{G.}~\bibnamefont{M\"{u}ller}},
  \bibinfo{author}{\bibfnamefont{H.}~\bibnamefont{Thomas}},
  \bibinfo{author}{\bibfnamefont{H.}~\bibnamefont{Beck}}, \bibnamefont{and}
  \bibinfo{author}{\bibfnamefont{J.~C.} \bibnamefont{Bonner}},
  \bibinfo{journal}{Phys. Rev. B} \textbf{\bibinfo{volume}{24}},
  \bibinfo{pages}{1429} (\bibinfo{year}{1981}).

\bibitem[{\citenamefont{Karbach et~al.}(1997)\citenamefont{Karbach, M\"{u}ller,
  Bougourzi, Fledderjohann, and M\"{u}tter}}]{Karbach_Bougourzi}
\bibinfo{author}{\bibfnamefont{M.}~\bibnamefont{Karbach}},
  \bibinfo{author}{\bibfnamefont{G.}~\bibnamefont{M\"{u}ller}},
  \bibinfo{author}{\bibfnamefont{A.~H.} \bibnamefont{Bougourzi}},
  \bibinfo{author}{\bibfnamefont{A.}~\bibnamefont{Fledderjohann}},
  \bibnamefont{and} \bibinfo{author}{\bibfnamefont{K.-H.}
  \bibnamefont{M\"{u}tter}}, \bibinfo{journal}{Phys. Rev. B}
  \textbf{\bibinfo{volume}{55}}, \bibinfo{pages}{12510} (\bibinfo{year}{1997}).

\bibitem[{\citenamefont{des Cloizeaux and Pearson}(1962)}]{Cloizeaux_Pearson}
\bibinfo{author}{\bibfnamefont{J.}~\bibnamefont{des Cloizeaux}}
  \bibnamefont{and} \bibinfo{author}{\bibfnamefont{J.~J.}
  \bibnamefont{Pearson}}, \bibinfo{journal}{Phys. Rev.}
  \textbf{\bibinfo{volume}{128}}, \bibinfo{pages}{2131} (\bibinfo{year}{1962}).

\bibitem[{\citenamefont{Baskaran et~al.}(1987)\citenamefont{Baskaran, Zou, ,
  and Anderson}}]{Baskaran}
\bibinfo{author}{\bibfnamefont{G.}~\bibnamefont{Baskaran}},
  \bibinfo{author}{\bibfnamefont{Z.}~\bibnamefont{Zou}}, , \bibnamefont{and}
  \bibinfo{author}{\bibfnamefont{P.~W.} \bibnamefont{Anderson}},
  \bibinfo{journal}{Solid State Commun.} \textbf{\bibinfo{volume}{63}},
  \bibinfo{pages}{973} (\bibinfo{year}{1987}).

\bibitem[{\citenamefont{Affleck and Marston}(1988)}]{Affleck_Marston}
\bibinfo{author}{\bibfnamefont{I.}~\bibnamefont{Affleck}} \bibnamefont{and}
  \bibinfo{author}{\bibfnamefont{J.~B.} \bibnamefont{Marston}},
  \bibinfo{journal}{Phys. Rev. B} \textbf{\bibinfo{volume}{37}},
  \bibinfo{pages}{R3774} (\bibinfo{year}{1988}).

\bibitem[{\citenamefont{Arovas and Auerbach}(1988)}]{Arovas}
\bibinfo{author}{\bibfnamefont{D.~P.} \bibnamefont{Arovas}} \bibnamefont{and}
  \bibinfo{author}{\bibfnamefont{A.}~\bibnamefont{Auerbach}},
  \bibinfo{journal}{Phys. Rev. B} \textbf{\bibinfo{volume}{38}},
  \bibinfo{pages}{316} (\bibinfo{year}{1988}).

\bibitem[{\citenamefont{Gagliano and Balseiro}(1987)}]{Gagliano_Balseiro}
\bibinfo{author}{\bibfnamefont{E.~R.} \bibnamefont{Gagliano}} \bibnamefont{and}
  \bibinfo{author}{\bibfnamefont{C.~A.} \bibnamefont{Balseiro}},
  \bibinfo{journal}{Phys. Rev. Lett.} \textbf{\bibinfo{volume}{59}},
  \bibinfo{pages}{2999} (\bibinfo{year}{1987}).

\end{thebibliography}
\end{document}